\documentclass[pre,a4paper,twocolumn,footinbib,showpacs]{revtex4}

\usepackage{amssymb}
\usepackage{amsmath}
\usepackage{amsbsy}
\usepackage{graphicx}

\begin{document}
\title{ Diffusion of a liquid nanoparticle on a disordered substrate. }

\author{Franck Celestini}
   \email{Franck.Celestini@unice.fr}
   \affiliation{Laboratoire de Physique de la Mati\`ere Condens\'ee
  ,UMR 6622, CNRS, Universit\'e de Nice Sophia-Antipolis,
   Parc Valrose  06108, 13384 Nice Cedex 2, France}

\date{\today}

\begin{abstract}
 We perform molecular dynamic simulations of liquid nanoparticles deposited on
 a disordered substrate. The motion of the nanoparticle is characterised by a 'stick and roll'
 diffusive process. Long simulation times ($\simeq \mu s$), analysis of mean 
square displacements and stacking time distribution functions demonstrate that the 
nanoparticle undergoes a normal diffusion in spite of long sticking times.
We propose a phenomenological model for  the size and temperature dependence
of the diffusion coefficient in which the activation energy scales as $N^{1/3}$.

\end{abstract}


\maketitle 

\section{Introduction}

The spatial organisation of nanoparticles deposited on a substrate 
strongly depends on the nature and the magnitude of their
mobilty during the growth process. For systems known as Volmer-Weber films \cite{volmer}
for which nanoparticles are isolated 
from each other, it is of importance to understand their diffusion on the substrate.
This diffusion is directly related to  the interaction between the
substrate and the nanoparticle so that depending on the system studied one can 
found slow or fast diffusing nanoparticles \cite{jensen,bardotti,wen}.
This has been illustrated by Deltour {\em et al.}  \cite{barrat} who demonstrated  that a  small change in the mismatch between
the lattice parameters of a Lennard-Jones cluster and the substrate induces an
important variation of the diffusion coefficient. More recently, a surprising increasing mobility
with increasing size of the nanoparticle has also been reported \cite{carrey}

From a more fundamental point of view it is  important to determine the nature of the nanoparticle
diffusion. The simplest method is to look at the time dependence of the mean
square displacement of the center of mass  : 
\begin{equation}
<R^2_{cm}(t)> = <({\bf r_{cm}}(t)- {\bf r_{cm}}(0))^2>  \propto  t^\gamma
\end{equation}
where $ {\bf r_{cm}}(t)$ is the position of the center of mass at time $t$. The exponent
$\gamma$ is equal to one for a normal diffusion while $\gamma > 1$ 
and $\gamma < 1$ correspond respectively to superdiffusive and subdiffusive
processes. Luedtke and Landman \cite{lued1,lued2} have performed simulations of  gold nanocrystals
adsorbed on graphite and exhibiting L\'evy Flights leading to mild superdiffusion 
($\gamma\simeq 1.1$). More recently and for a similar system Maruyama and Murakami \cite{maru}
found a crossover between superdiffusion and normal diffusion. These different 
diffusion regimes are influenced by the way the nanoparticle moves on the surface. Sliding of the
whole nanoparticle seems the most probable scenario but a rolling mechanism has also been 
observed for small clusters \cite{fan}.
 
\begin{figure}
\includegraphics[width=0.45\textwidth]{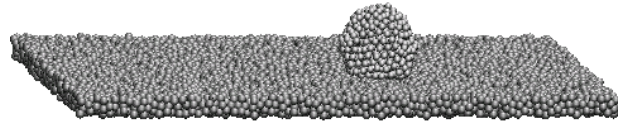}\\
\caption{Snapshot of a $N=555$ gold liquid nanoparticle deposited on the disordered substrate.
We measure a wetting angle  $\theta \simeq 140^\circ$ }
\label{figure1}
\end{figure}

To our knowledge and in contrast to the case of solid nanoparticles, a single study \cite{lewis} has been
devoted to the diffusion of liquid nanoparticles. The authors have reported a discontinuity in the diffusion
coefficient obtained for solid and liquid nanopaticles near the melting point.
In this paper we perform molecular
dynamics simulations of gold liquid nanoparticles deposited on a weakly disordered substrate.
 The first section  of this paper presents the simulation details
and in the second one we characterize the motion of the nanoparticle
on the substrate. We show that  it is made of sticking and moving events. The existence of a contact 
angle hysteresis is at the origin of the oscillations  around an equilibrium position. 
A visual inpection and the measure of the correlation between the velocity and the angular momentum 
evidence the tendency of the nanoparticle to experience a 'rolling like' motion to reach another equilibrium position.
In the  third section we examine the nature of the diffusion. Long simulation times are necessary to compute
reliable mean squared displacements and sticking time distribution functions. We show that the diffusion 
remains normal even for the largest nanoparticles having large sticking characteristic times. A 
phenomenological model is finally presented to account for the size and temperature dependence of the
diffusion coefficient. The main hypothesis of a size dependent activation energy is necessary to recover
the simulation results.

\section{Simulation procedure :}

 The nanoparticles considered in this study have  sizes 
ranging between $N=87$ and $N=555$ atoms. We use the Ercolessi glue potential \cite{furio}  to describe the 
 interactions between Au atoms with a cut-off distance of $3.9 \AA$ and a time step $\delta t=
 2.5$ $10^{-3} ps$. In this formalism the potential energy of the $N$ atoms is given by :
\begin{equation}
U=\frac{1}{2}\sum_{i,j=1}^N \Phi(r_{ij}) +\sum_{i=1}^N U(n_i)
\end{equation}
 The first term is  a classical pair interaction. In the second term,
 $n_i$ is the coordination of atom $i$ :
\begin{equation}
n_i=\sum_{j=1}^{N} \rho(r_{ij})
\end{equation}
 where $\rho(r_{ij})$ is a function of the interatomic distance $r_{ij}$.
 The energy function $U$ is the {\em glue term} associating an extra potential
 energy which is a function of the atom coordination.
\begin{figure}
\includegraphics[width=0.4\textwidth]{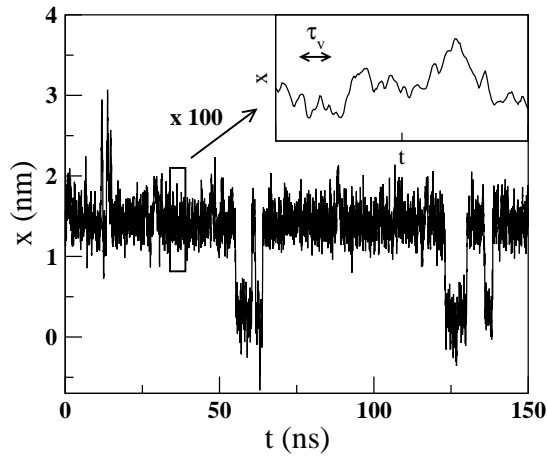}\\
\caption{ Time evolution of the $x$ coordinate of the center of mass for a particle of 
$555$ atoms. The sticking and moving events are clearly identified. Inset : during a sticking event 
the CM oscillates around an equilibrium position with a characteristic  vibrational period $\tau_v$ .
 }
\label{figure2}
\end{figure}

The substrate is a slab of dimension $170*170*10 \AA^3$  made of $15562$ atoms. The substrate
atomic positions come from the simulation of a liquid  rapidly quenched
below its melting temperature. We  obtain a substrate with a glass structure and
 periodically boundary conditions  applied in the $x$ and $y$ directions. 
 The substrate atoms  are chosen to be frozen in  order to increase the simulation time.
 We  therefore expect a quantitative change of the diffusion coefficient as compared
 to a more realistic simulation in which the substrate atoms can move \cite{lewis}. Nevertheless
this choice permits to reach simulation times of roughly $1 \mu s$ that, as will be discussed
below, are necessary to clearly identify the diffusion regime.
 The interaction between   substrate
 and Au atoms are chosen to be  of the Lennard jones type : 
 \begin{equation}
V_{ij} = 4 \epsilon [(\frac{\sigma}{r})^{12}-(\frac{\sigma}{r})^6]
 \end{equation}
where $r$ is the distance between the two atoms. We choose $\sigma=2.7 $ $\AA$, 
$\epsilon=255$ $ k_b $ and a cut-off distance
of $2.5$ $ \sigma$. As a consequence of this  weak  $\epsilon$ value, the liquid Au
nanoparticles are in a weakly wetting situation with a wetting angle roughly estimated to $\theta_e 
\simeq 140 ^ \circ$. A snapshot of the simulated system is given in Figure 1. where
we can see a $N=555$ liquid nanoparticle deposited on the disordered substrate.
 The equations of motion  are integrated using the Verlet algorithm and the 
temperature of the nanoparticle is fixed using the classical velocities rescaling
procedure \cite{allen}. We choose  $T=800 K$ below the bulk melting point
($T_m \simeq 1380 K$) but still above the solidification temperature. The liquid state
of the nanoparticles is evidenced by the atomic mean squared displacements calculated
in the frame of the center of mass. The values obtained for the atomic diffusion 
are in good agreement with previous ones \cite{lewis}. For example, for a particle of $249$ atoms the diffusion coefficient is estimated to be $0.09 \AA ^2 ps^{-1}$.
  In a first run of roughly $10^7$ MD steps the liquid nanoparticle is deposited on the 
 substrate and reach its equilibrium shape. A second long run of $10^9$ MD steps is then
performed. It permits to record the position of the center of mass and to compute different
quantities like the probability density function of the sticking events. We checked that the same 
results are obtained if the second run is performed under microcanonical conditions. The
velocity rescaling procedure has therefore no influence on the nanoparticle motion.
  
\section{Analysis of the liquid drop motion}
  
\begin{figure}[b]
\includegraphics[width=0.35\textwidth]{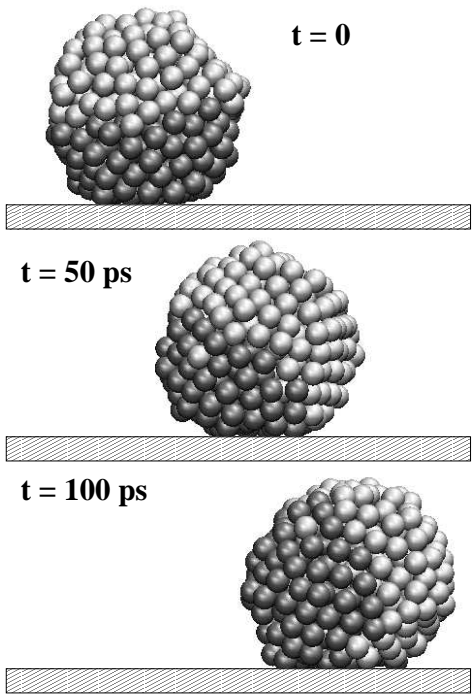}\\
\caption{ Snapshots of a moving nanoparticle taken at times separeted by $50 ps$. The atoms 
initially belonging to the lower half of the particle are grey colored to evidence the 'rolling like' move.
 }
\label{figure3}
\end{figure}

The trajectories of the center of mass (CM) show that the diffusion process is composed of two
distincts events : sticking events in which the CM oscillates around an
equilibrium position and moving events in which the CM reaches another equilibrium position. 
This is illustrated in fig. 2. where we plot the time evolution of the CM x coordinates $x_{cm}(t)$
 recorded during $20 ns$ for a particle of $555$ atoms. For this large particle, the sticking and moving
events are clearly identified. In inset we represent an hundredfold time expansion of $x_{cm}(t)$
for the same system. During a sticking event the CM oscillates around an equilibrium position 
with a characteristic  vibrational period $\tau_v$. A Fourier transform of $x_{cm}(t)$ and $y_{cm}(t)$ permits
to evaluate $\tau_v$. These results will be presented elsewhere but as a main result the 
vibrational period $\tau_v$  is found to scale with the square root of the particle size. 
Indeed because of the hysteresis contact angle \cite{deg} the liquid drop is at rest on the
substrate and is exited by the thermal noise. The measured  $\tau_v$  corresponds to the
eigen vibrational period due to the elastic deformation of the drop :
\begin{equation}
\tau_v \propto (\frac{\rho}{\gamma_{lv}})^{1/2} N^{1/2}
\end{equation}
where $\rho$ is the density and $\gamma_{lv}$ the liquid-vapor surface tension.  
 This period is similar to the one found for the shape oscillations of a free liquid drop \cite{hydro}.
When the particle leaves its equilibrium position, we observe a 'rolling like' move (note that
the liquid drop is also changing its shape during the move) 
of the liquid drop  on the substrate. In fig. $3$  we represent three snapshots of
a moving particle at times separeted by $50$ $ps$. The atoms initially belonging to the lower half
of the particle are grey colored to evidence the 'rolling like' motion. To quantify this visual inspection we
calculate the correlation between the  components of the CM velocity ($V_x$ and $V_y$) and the orbital
moments ($J_x$ and $J_y$). As usually the correlation coefficient $C_{ij}$ between the components 
$i$ and $j$ are defined as : 
\begin{equation}
C_{ij} = \frac{<V_iJ_j>-<V_i><J_j>}{\sigma_{V_i}\sigma_{J_j}}
\end{equation}

We report on Table I the different $C_{ij}$ coefficients computed for a $N=321$ particle during a run of
$150 ns$. The correlation is not significative between $V_x$ and $J_x$ and between $V_y$ and $J_y$.
Conversely and as 
expected for a 'rolling like' move, a correlation is found for the cross components of the velocity and the
orbital moment : $C_{xy} \simeq - C_{yx} \ne 0$. 

\vspace{0.5cm}
\begin{center}
\begin{tabular}{| p{1.cm}|p{1.5cm}|p{1.5cm}|}
\hline
    $\ C_{ij}$ & \ x  &  \ y  \\
\hline
$ \ x$ & \ $-0.002$ & \ $0.120$ \\
\hline
$ \ y$ & \ $-0.137$ & \ $0.007$ \\
\hline   
\end{tabular}
\vspace{0.5cm}

\end{center}

\begin{figure}[t]
\includegraphics[width=0.45\textwidth]{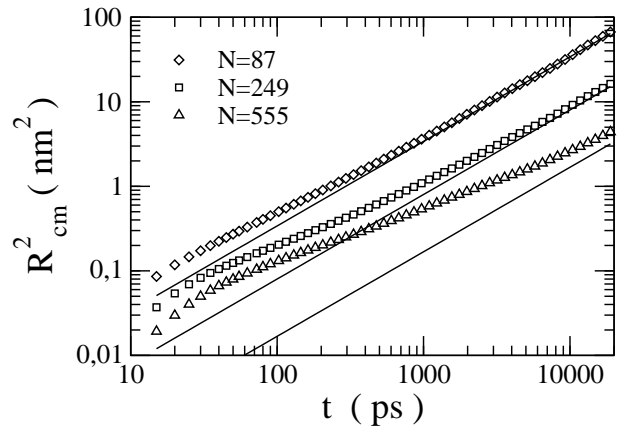}\\
\caption{ Mean square displacement $R^2_{cm}(t)$  for particles with sizes $N=87$, $249$ and
$555$. The straigth lines correspond to the asymptotic linear regimes : $R^2_{cm}(t) \propto 4 D t$.
 }
\label{figure4}
\vspace{0.5cm}
\end{figure}
Note finally that these  correlations coefficients are surestimated because they also characterize the 
vibrations around the equilibrium position. Therefore and since the radius of our particles should be
of the same order than the slipping length \cite{barslip,nature}, we cannot rule out possible sliding
moves of the nanoparticle.

\section{Size and temperature dependence of the diffusion coefficient}

We now present the results obtained for the mean square displacement $R^2_{cm}(t)$ of particles
with sizes ranging from $N=87$ to $N=555$ atoms. We plot in figure 4. the statistically reliable
values of $R^2_{cm}(t)$ ( $t<20$ $ns $) computed for a run length of roughly $1\mu s$.
For the smallest particle ($N=87$) the regime of normal diffusion is 
clearly indentifiable with $R^2(t) \propto t$. As the particle size increases the time to
reach a linear regime increases. For the largest particle considered in this study ($N=555$)
the linear regime is not reached and the conclusion from the $R^2(t)$ curve is not clear : one 
could think at a subdiffusive regime with a fitted exponent $\gamma \simeq 0.72$ or alternatively 
that the simulation time is not long enough to evidence a normal diffusion regime. 

\begin{figure}[t]
\vspace{1cm}
\includegraphics[height=0.25\textheight]{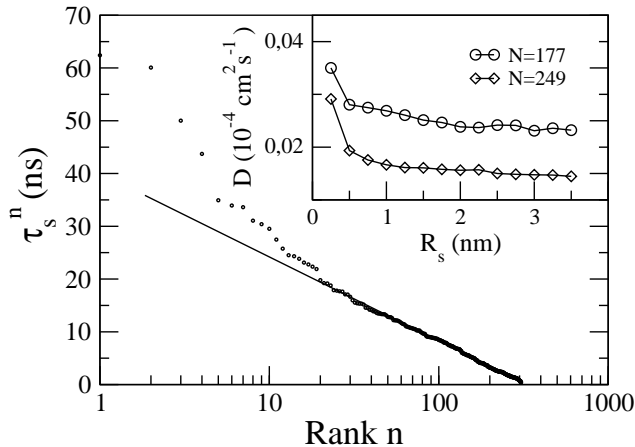}\\
\caption{ Rank ordering of the sticking times for a particle of $249$ atoms ($R_s=25\AA$). The slope
of the straight line corresponds to the mean sticking time of the exponential distribution. Inset : the
diffusion constants  $D(R_s)$ tend to a constant for sufficiently large $R_s$ values.
 }
\label{figure5}
\end{figure}

In order to reach a  conclusion about the nature of the diffusion for the largest particles, we look at 
the probability density function (PDF) $\rho_R(\tau)$ of the sticking events. Indeed we know \cite{bouch}
that the diffusion nature depends on the form of $\rho_R(\tau)$ :  for a PDF that has a mean 
value (like an exponential one) the diffusion is known to be normal while  for a power law PDF 
( $ \propto \tau ^{-(1+\mu)}$ with $ 0<\mu<2)$ ) we expect a subdiffusive behavior. 
The time $\tau_s$ of a sticking event is defined as the necessary time for the CM to cross a 
circle of radius $R_s$ centered on its initial position. We compute the PDF for  different particle
sizes and different values of $R_s$. Since the statistic on sticking times is rather poor especially
for the largest systems, the nature of the PDF is determined using the  rank ordering method \cite{didier}. 
For an exponential distribution we expect that the the $n-th$ largest value of $\tau_s$, $\tau_s^n$
scale as :
\begin{equation}
\tau_s^n = - \bar\tau_s log(n/N)
\end{equation}
where $N$ is the total number of observed sticking events and $\bar\tau_s$ the mean  sticking time.
As can be seen in figure 5. for a particle
of $249$ atoms and $R_s=25 \AA$ the function $\tau_s^n$ is clearly logarithmic in $n$. Whatever
the particle size considered, for sufficiently large $R_s$ values
the associated PDF is exponential and we can measure a mean  sticking time $\bar \tau_s$. This analysis 
demonstrates that the diffusion remains normal even for the largest particles. The apparent subdiffusive 
behavior observed in the  $R^2_{cm}(t)$ curves is  due to the large value of the mean sticking 
time as compared to the simulation time. We also compute the mean square displacement $\bar r^2$ 
between two sticking events and finally estimate the diffusion coefficient to be $ D(R_s) =
\bar r^2 /4  \bar \tau_s$. We can see on the inset of figure 5. that $D(R_s)$ tends to  a constant for
sufficiently large $R_s$ values. We  establish here a first method to measure the diffusion
coefficient.

\begin{figure}[b]
\vspace{1cm}
\includegraphics[width=0.45\textwidth]{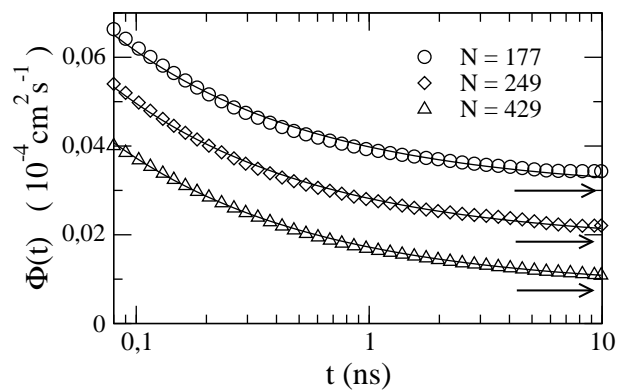}\\
\caption{ Assymptotic analysis of $\Phi(t)=R^2_{cm}(t)/4t$ for nanoparticles with $N=177, 249 and 429$.
Full lines are best fits to the simulated data and the arrows indicate the extrapolated values of
$D$.
 }
\label{figure6}
\end{figure}

Another possible way to extract $D(N)$ from the simulation data is to look at the
asymptotyc behavior of $R^2_{cm}(t)$. We have demonstrate above that the diffusion remains normal
whatever the particle size so that the quantity $\Phi(t)=R^2_{cm}(t)/4t$ should be of the form :
\begin{equation}
\Phi(t) =  D  + \alpha t^\gamma      
\end{equation}
for large $t$ values with an exponent $\gamma<0$ to ensure the
asymptotic normal diffusion regime. Best fits of our data
to this form of $\Phi(t)$ give values of $\gamma$ close to $-0.5$ with a small dispersion around
this mean value ($\sigma_\gamma  \simeq 0.05$). This exponent is the same than the one predicted 
for the model of continuous time random walk \cite{bouch} with a
sticking time distributed according to a power law $\rho(\tau) \propto \tau ^{-(1+\mu)}$ with
$ \mu<3$ (normal diffusion regime). We therefore extract the diffusion 
coefficients $D(N)$ through  best fits of simulated $\Phi(t)$ values to equation $8$ (the value
of $\gamma$ is fixed to $0.5$ for the fitting procedure).
We represent in figure 6. the simulated and fitted $\Phi(t)$ values for systems with $N=177$, $249$ and $429$
atoms. The arrows indicate the values of the diffusion coefficient extracted from this asymptotic 
analysis.

\begin{figure}[b]
\vspace{1cm}
\includegraphics[width=0.45\textwidth]{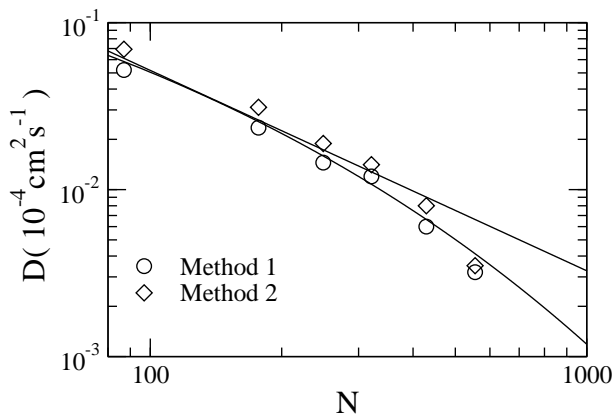}\\
\caption{ Diffusion coefficients as a function of the system size. We represent the value obtained
by the two methods. The two straight lines correspond to the best fits to a power law and to a law
in which the activation energy is a function of the system size.
 }
\label{figure7}
\end{figure}

We represent in figure 7. the diffusion coefficients obtained by the analysis of the sticking times PDF and by
the second method  described just above as a function of the system size. 
A first fit of these data to a power law $D(N) \propto N^{\gamma}$ leads to a value $\gamma = -1.3 \pm 0.1$.
 The agreement is satisfactory for the smallest particles but we observe a deviation for the largest
 $N$ values. The power law 
 apparently tends to overestimate the value of $D$.
 For a LJ cristalline nanoparticle on a solid substrate \cite{barrat}  
 the exponent is $\gamma = 2/3$ for a large mismatch parameter. This exponent has been recently confirmed
for gold supported nanoparticles \cite{antonov}. In the case of quasi epitaxy the
 exponent increases apparently up to $\simeq 1.4$ \cite{barrat}. This latter value is close to the one  
 found in the present study for a liquid nanoparticle. 
 In both cases, a  critical applied force exists, below which the system cannot move. In our case
 this force is related to the 
 wetting contact angle hysteresis. To our knowledge no theoretical prediction of this large exponent
 value exists. Note finally that the diffusion coefficients calculated with the fitted power law for large
 particle sizes ( for example $D \simeq 4$  $nm^2 s^{-1} $ for a particle of radius $\simeq 100 nm$ ) 
 are definitely too large to be consistent with experimental observations \cite{richard1,richard2} 
 of  completely sticked liquid nanoparticles. This suggests that the power law is no longer vadid for
 large particles.
 
 We therefore propose a second phenomenological scaling law  of the form :
 $D(N) \propto exp(-\alpha N^{1/3}/kT)$. In this expression the activation energy of the diffusive process is
a function of $N$. To escape from its equilibrium position the liquid drop is elastically deformed with 
advancing and receding wetting angles differents of the equilibrium one. The exponent $1/3$ in the proposed
scaling law relates the elastic deformation caused by the wetting angle hysteresis and the parameter
$\alpha$ is therefore  proportional to $\gamma_{lv}$. We can see on fig. 7. that this scaling law reasonably fit
our data on the whole range of particle sizes with a free parameter $\alpha = 0.051 eVatoms^{-1/3}$. 
As done just above we can estimate the value of $D$ for large particles. This time the result found is in agreement
with the experimental observation \cite{richard1,richard2}. Note that we have made the hypothesis that the main $N$ dependence
is in the exponential. Nevertheless a size and temperature dependent prefactor should also exists so that
a more general  expression  should be of the form :
\begin{equation}
D(N,T) \propto  e^{-\Delta E_i/kT}  f(N) e^{-\Delta E_c(N)/kT}
\end{equation}
The first exponential term contains the activation energy of individual atoms. In the present case
of a liquid nanoparticle this activation energy is related to the atomic
diffusion within the liquid. The two other terms have been discussed above, they are reflecting the
collective diffusion of the nanoparticle. The form of the function $f(N)$ depends on the nature of 
the dissipation and could be evaluated in the same way than for a millimetric drop \cite{pom}. 

\begin{figure}[b]
\vspace{1cm}
\includegraphics[width=0.45\textwidth]{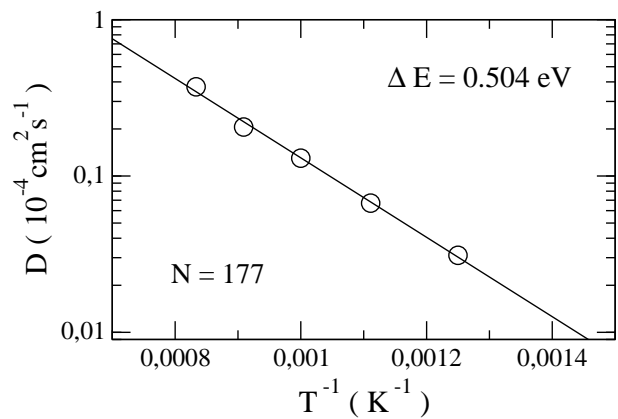}\\
\caption{ Diffusion coefficients as a function of temperature for a $N=177$ particle. The best fit to an arhenius law gives 
an activation energy of $0.504 eV$.
 }
\label{figure8}
\end{figure}

 To confirm this exponential relation between the diffusion coefficient and the particle size
we finally look at the temperature dependence of $D$ for a particle of $177$ atoms.
 The diffusion coefficients are computed for temperatures between $700$ and $1200 K$. Data are 
 well fitted by an arhenius law $D \propto exp(-\Delta E/kT)$ and give an activation energy 
$\Delta E= 0.504 eV$ (Fig. 8.). This value is greater than the one obtained from the previous analysis
of the $D(N)$ data. Using the fitted $\alpha$ value we indeed found $\Delta E_c(N=177) = 0.286 eV$.
Equation $9.$  predicts that the overall activation energy $\Delta E$ should be the sum of
$\Delta E_i$ and $\Delta E_c$. We then perform simulation to measure the atomic diffusion $D_i$ within
the liquid. The values $D_i(T)$ are extracted from mean square displacement  calculated in the
frame of the CM for different temperatures. An arhenius law fits the data and gives the value of
the individual activation energy : $\Delta E_i = 0.196 eV$. Finally we found  $\Delta E_c(N=177) +
\Delta E_i = 0.482 eV$ that compares rather well with $\Delta E= 0.504 eV$ and seems to confirm the
phenomenological form proposed for  $D(N,T)$ in Equ. $9$.

\section{Conclusions and perspectives}

To summarize, we have performed molecular dynamic simulations of a liquid nanoparticle diffusing
on a weakly disordered substrate. For the  weakly wetting condition considered here, the nanoparticle experiences
a 'rolling like' motion from an equilibrium position to another one. The sticking time
distribution function is shown to be exponential even for the largest nanoparticles. As
a consequence the diffusion is normal with a mean square displacement asymptotically
linear in $t$. The diffusion coefficients are reported as a function of size and temperature. A 
phenomenologicalexpression for $D(N,T)$ is proposed. The main hypothesis of an activation energy
scaling with $N^{1/3}$ is necessary to fit the simulation results.
	 
In spite of the difficulty of a direct observation of the nanoparticle diffusion, we
hope this paper will motivate further experimental studies. Finally, a work is currently in progress to 
fully characterize the vibrational modes of the sticked liquid nanoparticle. 

\begin{acknowledgments}
 I would like to thank  A. ten Bosch, R. Kofman and L. Lobry for fruitfull discussions.
\end{acknowledgments}


\begin{references}

\bibitem{volmer} M. Zinke-Allmang, L. C. Feldman and M. Grabow,
Surf. Sci. Rep. {\bf 16}, 377 (1992).

\bibitem{jensen} P. Jensen, Rev. Mod. Phys. {\bf 71}, 1695 (1999).

\bibitem{bardotti}
L. Bardotti, P. Jensen, A. Hoareau, M. Treilleux and 
B. Cabaud, Phys. Rev. Lett. {\bf 74}, 4694 (1995).

\bibitem{wen} J. M. Wen, S. L. Chang, J. W. Burnett, J. W. Evans and P. A. Thiel,
 Phys. Rev. Lett. {\bf 73}, 2591 (1994).

\bibitem{barrat}
P. Deltour, J. L. Barrat and P. Jensen, Phys. Rev. Lett. {\bf 78}, 4597 (1997).

\bibitem{carrey} J. Carrey, J. L. Maurice, F. Petroff and A. Vaures,
Phys. Rev. Lett.  {\bf 86}, 4600 (2001).

\bibitem{lued1}
W. D. Luedtke and U. Landman, J. Phys. Chem. {\bf 100}, 13323 (1996).

\bibitem{lued2}
W. D. Luedtke and U. Landman, Phys. Rev. Lett. {\bf 82}, 3835 (1999).

\bibitem{maru} Y. Maruyama and J. Murakami,  Phys. Rev. B {\bf 67}, 85406 (2003).

\bibitem{fan} W. Fan, X. G. Gong and W. M. Lau,  Phys. Rev. B {\bf 60}, 10727 (1999).

\bibitem{furio}
F. Ercolessi, M. Parrinello and E. Tosatti, Philos. Mag. A {\bf 58}, 213 (1988).

\bibitem{lewis} L. lewis, P. Jensen, N. Combe and J. L. Barrat,
Phys. Rev. B {\bf 61}, 16084 (2000).

\bibitem{allen} M. P. Allen and D. J. Tildesley, {\em Computer Simulation of Liquids}
(Claredon Press, Oxford, 1987).

\bibitem{deg} J. F. Joanny and P. G. de Gennes, 
J. Chem. Phys. {\bf 81}, 552 (1984).

\bibitem{hydro} S. Chandrasekhar, {\em Hydrodynamic and Hydromagnetic Stability}
(Oxford University Press, London, 1961).

\bibitem{barslip}  J. L. Barrat and L. Bocquet,
Phys. Rev. Lett. {\bf 82}, 4671 (1999).

\bibitem{nature} P.A. Thomson and S. M. Troian, Nature (London) {\bf 389}, 360 (1997).

\bibitem{bouch}
J. P. Bouchaud and A. Georges, Phys. Rep. {\bf 195}, 127 (1990).

\bibitem{didier} D. Sornette, {\em Critical Phenoma In Natural Sciences}
(Springer Verlag, Berlin, 2000).

\bibitem{antonov} V. N. Antonov, J. S. Palmer, A. S. Bhatti and J. H. Weaver,
Phys. Rev. B  {\bf 68}, 205418 (2003).

\bibitem{richard1} E. Sondergard, R. Kofman, P. Cheyssac and A. Stella,
Surf. Sci. {\bf 364}, 467 (1996).


\bibitem{richard2} L. Haderbache, R. Garrigos, R. Kofman, E. Sondergard and P. Cheyssac,
Surf. Sci. Lett. {\bf 410}, L748 (1998).

\bibitem{pom} L. Mahadevan and Y. Pomeau,
Phys. Fluids  {\bf 11}, 2449 (1999).

\end{references}
\end{document}